\documentclass[preprint,aps]{revtex4}
\usepackage{graphicx}
\usepackage{epsfig}
\usepackage{epstopdf}
\usepackage{dcolumn}% Align table columns on decimal point

\usepackage{bm}% bold math
\usepackage{amsmath}% AMS math commands
\usepackage{amssymb}
\usepackage{placeins} %to use \FloatBarrier command

\def\be{\begin{equation}}
\def\ee{\end{equation}}
\def\bea{\begin{eqnarray}}
\def\eea{\end{eqnarray}}

\begin{document}

\title{Transverse Momentum and Transverse Momentum Distributions in the MIT Bag Model}

%\pacs{12.38.Aw, 12.39.Ba, 14.20.Dh }
%\keywords      {Quark distributions, Quark model, Nucleon}

\author{A. I. Signal}
\email{a.i.signal@massey.ac.nz}{
\affiliation{School of Fundamental Sciences PN461 \\ Massey University \\ Palmerston North 4442 \\ New Zealand}

\author{F. G. Cao}
\email{f.g.cao@massey.ac.nz}{
\affiliation{School of Fundamental Sciences PN461 \\ Massey University \\ Palmerston North 4442 \\ New Zealand}

\vskip0.5cm

\begin{abstract}

The typical transverse momentum of a quark in the proton is a basic property of any QCD based model of nucleon structure. 
However, calculations in phenomenological models typically give rather small values of transverse momenta, 
which are difficult to reconcile with the larger values observed in high energy experiments such as Drell-Yan reactions
and Semi-inclusive deep inelastic scattering. 
In this letter we calculate the leading twist transverse momentum dependent distribution functions (TMDs) using 
a generalization of the Adelaide group's relativistic formalism that has previously given good fits to the parton distributions. 
This enables us to examine the $k_{T}$ dependence of the TMDs in detail, and determine typical widths of these distributions.
These are found to be significantly larger than those of previous calculations.
We then use TMD factorization in order to evolve these distributions up to experimental scales where we can compare with data on 
$\langle k_{T} \rangle$ and $\langle k^{2}_{T} \rangle$. 
Our distributions agree well with this data. \\

PACS: 12.38.Aw, 12.39.Ba, 14.20.Dh

\end{abstract}

\maketitle

%%%%%%%%%%%%%%%%%%%%%%%%%%%%%%%%%%%%%%%%%%%%
%% MAINMATTER
%%%%%%%%%%%%%%%%%%%%%%%%%%%%%%%%%%%%%%%%%%%%

%\section{Introduction}

%Transverse momentum of quarks
%Natural scale for models?
%Nucleon size around 1 fm implies average $p_{T}$ around 0.4 GeV
%The typical transverse momentum of a quark in the nucleon is a basic property of any QCD-based model of nucleon structure. 
%For instance, the average transverse momentum $\langle p_{T} \rangle$ is thought to set a `natural' scale for such models, 
%and that a nucleon size around 1 fm implies $\langle p_{T} \rangle \approx 0.4$ GeV \cite{TW01}. %references?
%However, this observable has been surprisingly difficult to determine, either from experiment or model calculation.

Transverse momentum dependent distributions (TMDs) allow us to investigate transverse momenta in the nucleon and many phenomena that depend upon transverse momenta.
For example, at leading twist TMDs can be used to describe Drell-Yan (DY) reactions \cite{Collins02}, 
Semi-inclusive deep inelastic scattering (SIDIS) \cite{Kotzinian95, Hermes13, Compass13} 
and hadron production in $e^{+}$-$e^{-}$ annihilation \cite{Collins93, Belle06, Belle11}. %references
Hence there has been much effort in recent times to calculate TMDs in various phenomenological models \cite{AESY10, PCB08, SZM09, EPTZ09, BCR08, MBCT12, Wakam09}. %references Schweitzer etc
As well as giving insight into experimental observables, model TMDs can provide new information about non-perturbative properties of the nucleon and other hadrons.
For instance, the Fourier transform conjugate variable to transverse momentum is the impact parameter $\mathbf{b}_T$, 
so taking the transform of a TMD gives the quark distribution in impact parameter space, and would give insight into possible breaking of spherical and axial symmetry 
in the quark wavefunction \cite{Burk00, Burk02, MIller07}. 

Previous calculations of TMDs using the bag model (and other models) have generally not had correct support, because momentum conservation has not been 
enforced in the scattering calculation \cite{AESY10, Yuan03, AESY08}. 
In Deep Inelastic Scattering (DIS) this leads to problems with interpreting the calculated parton distribution functions (PDFs), for instance non-zero distributions at $x = 1$ and beyond, 
and negative anti-quark distributions  \cite{ST89, SST91, SMST92}. 
There is no reason to believe these problems will be circumvented in calculations of TMDs with incorrect support.
Longitudinal momentum is constrained by the scattering dynamics, leading to the correct support for the plus component of momentum 
($k^{+} = (k^{0} + k^{3})/\sqrt{2}$) of the struck quark. 
Transverse momentum of the struck quark is also restricted $k_{T}^{2} \leq Q^2 (1-x) / 4x$ \cite{GSS07}, 
though this is usually ignored in the Bjorken limit. 
Nevertheless, the transverse and longitudinal momenta are not independent, and care needs to be taken when investigating distributions that depend on 
both momentum components.
There is a subtle distinction here from the case of light-cone wavefunctions, where the transverse and plus components of quark momentum are independent because the 
struck quark and the recoil/spectator state are both on-shell \cite{LB80, PB85}. 

A general TMD is described in terms of light-cone correlators (where we ignore the QCD gauge link between the quark operators) \cite{Collins11}
\bea
\phi \left(x, k_T \right)_{ij} = \int \frac{d^2 z_T}{(2 \pi )^3}dz^{-} e^{ikz} \left\langle \text{P}(P,S) \left| \bar{\psi }_{j }(0) \psi _{i}(z) \right| \text{P}(P,S)\right\rangle |_{z^+=0,k^+= xP^+} .
\label{eq:pwf}
\eea

Here we follow the notation and conventions of reference \cite{AESY10}, except we use $k_{i}$ and $p_{ni}$ to refer to components of quark and recoil state momenta respectively. 
We can insert a complete set of states $\sum_{n} |n \rangle \langle n |$ between the quark operators, then use translation invariance to express all spatial dependence in the exponential. 
The integrals over $z^{-}$ and $z_{T}$ will give delta functions 
$\delta \left(p_{n}^{+} - (1-x)P^{+}\right)$ and 
$\delta^{(2)} \left( \mathbf{p}_{nT} + \mathbf{k}_{T} \right)$, 
which express momentum conservation on the light-cone and in the transverse plane respectively.
The delta function on the light-cone constrains the plus component of momentum of the struck quark $k^{+} = xP^{+}$, 
leading to the TMDs only having support on the interval $0 \leq x < 1$. 
Also, this delta function constrains both the longitudinal and transverse components of the momentum $\mathbf{p}_{n}$ of the recoil state:
\bea
p_{n3} & = & M(1-x) - \sqrt{M_{n}^{2} + \mathbf{p}_{n}^{2}} \\
p_{nT}^{2} & = & 2M(1-x) \sqrt{M_{n}^{2} + \mathbf{p}_{n}^{2}} - (1-x)^{2}M^{2} - M_{n}^{2} 
\eea

where we are working in the LAB frame ($\mathbf{P} = 0$) and $M_{n}$ is the mass of the recoil state. 
In the large $x$ ($ p_{n} = |\mathbf{p}_{n}| \rightarrow \infty$) limit we have 
\bea
\frac{p_{n3}^2}{p_{n}^{2}} & \rightarrow & 1 + {\cal O} \left ( \frac{1}{p_{n}} \right) \\
\frac{p_{nT}^{2}}{p_{n}^{2}} & \rightarrow & {\cal O} \left ( \frac{1}{p_{n}} \right).
\label{eq:pntlim}
\eea

The consequence of this is that as the recoil momentum becomes large, it is dominated by the longitudinal component $p_{n3}$ 
and large values of transverse momentum $p_{nT}$ are not kinematically accessible. 
This means that integrals over transverse momentum, which will be required to calculate PDFs, moments and Fourier transforms, must have a large momentum cut-off. 
Alternatively, we can use the magnitude of the recoil momentum $p_{n}$ as the integration variable, subject to 
\bea
p_{n} \geq p_{\mathrm{min}} = \left| \frac{M^{2}(1-x)^{2} - M_{n}^{2}}{2M(1-x)} \right|
\eea

which comes from the requirement that $p_{nT}^{2}$ is positive definite. 
This change of variable gives expressions for the PDFs that agree with those of the Adelaide group \cite{SST91, Signal97}.

To obtain momentum eigenstates $| P \rangle$ and $| {\bf p}_{n} \rangle$ we use a Peierls-Yoccoz projection \cite{PY57} of MIT bag states. 
Using the MIT bag model wavefunction the general TMD can now be written as 
\bea
\phi(x,\ k_T) & = & 
N_{\phi } \frac{1}{P^+} \frac{| \varphi_{2} ({\bf p}_n) |^{2}}{| \varphi _{3}(0) |^{2}} f_{\phi}(t_{0}({\bf p}_{n} ), t_{1}({\bf p}_{n})) \delta(x - (1- \frac{{\bf p}_{n}^{+}}{P^{+}})) \delta ({\bf p}_{n_{T}} +{\bf k}_{T})
\eea

where $N_{\phi }$ is an appropriate spin-flavour matrix element \cite{SST91}, 
$ |\varphi_{2,3} ({\bf p}) |^{2}$ are the Fourier transforms of the 2 and 3 quark Hill-Wheeler overlap of the bag wavefunction, 
and $f_{\phi}$ is the required combination of the momentum space bag wavefunctions, given by equations (20) - (33) in reference \cite{AESY10}.
The bag wavefunction in momentum space is 
\bea
\psi(\kappa) = N \left(
\begin{array}{c}
t_0(\kappa) \\ i \sigma .\hat{\kappa} t_1(\kappa) \\
\end{array}
\right) \chi_{m}
\label{eq:bwf}
\eea

with
\bea
t_{0}(\kappa) & = & \frac{\omega^{2}}{\kappa(\kappa^{2} - \omega^{2})} \left( \omega \cos(\omega) \sin(\kappa) - \kappa \cos(\kappa) \sin(\omega) \right) \\
t_{1} (\kappa) & = & \frac{\omega^{3}}{\kappa^{2} - \omega^{2}} \left( \omega j_{0}(\omega)  j_{1}(\kappa) - \kappa j_{0}(\kappa) j_{1}(\omega) \right) 
\eea

where $\kappa = kR$, $\omega = 2.04 \ldots$ is the ground state energy eigenvalue and $j_{0.1}$ are spherical Bessel functions.

At leading twist, there are six T-even TMDs. 
However, these are not all independent in quark models (and the MIT bag model in particular), as these models generally do not have gauge field degrees of freedom, 
and Lorentz invariance provides further constraints \cite{AESY10, LP11, LP12}.
We choose to investigate the distributions $f_{1}(x,\ k_T)$, $g_{1}(x,\ k_T)$, $h_{1}(x,\ k_T)$ and $h_{1T}^{\perp}(x,\ k_T)$. 
The integrals over $k_T$ of the first three yield the familiar unpolarized, polarized and transversity PDFs $f_{1}(x)$, $g_{1}(x)$ and $h_{1}(x)$ respectively, 
while the fourth (pretzelosity) distribution is related to the quark orbital angular momentum in the model:
\bea
L^{3}_{q} = \int dx \: l^{3}_{q}(x) = - \int dx \; d^{2} \mathbf{k}_T  \; \frac{k_{T}^{2}}{2M^{2}} h_{1T}^{\perp}(x,\ k_T),
\eea
where we have introduced the orbital angular momentum density $l^{3}_{q}(x)$.

In figure 1 we show 3d plots of $f_{1}(x,\ k_T)$, $g_{1}(x,\ k_T)$, $h_{1}(x,\ k_T)$ and $h_{1T}^{\perp}(x,\ k_T)$ for a bag radius $R$ of 0.8 fm and a recoil mass $M_{n} = 0.75 M$. 
Of particular note is that $f_{1}(x,\ k_T)$, $g_{1}(x,\ k_T)$ and $h_{1}(x,\ k_T)$ all peak at $k_T = 0$ and are small for $k_T > 1$ GeV, 
while $h_{1T}^{\perp}(x,\ k_T)$ shows interesting structure and a definite minimum at $k_T > 0$ for $x < 0.5$.
In contrast to the results of reference \cite{AESY10}, we see that these distributions only have support on $0 \leq x < 1$, and are normalizable on this interval 
so that no renormalization procedure is necessary to calculate moments or the Fourier transforms of these distributions. 
Also, these distributions are well-behaved as $x \rightarrow 1$ and go to zero in this limit, in accordance with the behaviour of ${\bf p}_{n_{T}}$ given in equation (\ref{eq:pntlim}). 

Integrating our PDFs over the interval $0 \leq x \leq 1$ will yield sum rules, as shown in table 1. 
In this approach, the number and spin sum rules, calculated from $f_{1}$ and $\frac{1}{2}g_{1} + l^{3}_{q}$ respectively, 
do not  give the expected quark model values of $1$ and $\frac{1}{2}$ respectively.
This occurs because only intermediate states with 2 quarks have been considered \cite{ST89, SST91},
and other intermediate states such as $ | qqqg \rangle $ and $ | qqqq \bar{q} \rangle$ have not been added to the sum over all intermediate states. 
The tensor charge, given by the integral over the transversity PDF $h_{1}^{q}$, is compatible with the recent determination of \cite{RB18}, and will be investigated in further work.

We note that our TMDs automatically the satisfy the relation for the pretzelosity distribution \cite{AESY10}
\bea
h_{1}(x,\ k_T) - g_{1}(x,\ k_T) = \frac{k_{T}^{2}}{2M^{2}} h_{1T}(x,\ k_T)
\eea
as the bag model obeys the conditions found in references \cite{LP11, LP12}.

\begin{table}
\begin{center}
\begin{tabular}{|c|c|c|c|c|} \hline
$\int dx \, f_{1}$   & $\int dx \, g_{1}$ & $\int dx \, h_{1}$ & $\int dx \, l^{3}_{q}$ & $\int dx \, (\frac{1}{2}g_{1} + l^{3}_{q})$ \\  \hline
0.78 & 0.67 & 0.73 & 0.013 & 0.35 \\ \hline
\end{tabular}
\end{center}
\caption{Sum rules for twist 2 flavour independent PDFs for bag radius $R = 0.8$ fm. }
\end{table}

We find that the transverse momentum dependences of $f_{1}(x,\ k_T)$, $g_{1}(x,\ k_T)$ and $h_{1}(x,\ k_T)$ are well-fitted by Gaussian distributions of the form 
\bea
\phi \left(x, k_T \right) = \phi(x) \frac{1}{\pi \langle k^{2}_{T}(x) \rangle_{G}} \exp \left( \frac{-k^{2}_{T}}{\langle k^{2}_{T}(x) \rangle_{G}} \right)
\eea

with the Gaussian width $\langle k^{2}_{T}(x) \rangle_{G}$ showing some $x$ dependence, as would be expected from kinematic arguments \cite{CHS77, DS77}. 
%We note that the experimental data from Hermes and Compass is obtained using the assumption of a Gaussian width that is independent of $x$ \cite{Hermes13, Compass13, ABGMP14}.
%A more general analysis of the data, without invoking the Gaussian ansatz, may show some $x$ dependence of these widths. 

In the bag model, flavour dependence is introduced through the colour hyperfine interaction. 
In this work we use an exact approach using hyperfine eigenfunctions, which raise the degeneracy of the masses of the singlet and triplet recoil states \cite{CT88, SST91}. 
This has been criticised as being inconsistent with the Pauli exclusion principle \cite{Isgur99}, however, careful consideration of the hyperfine eigenfunctions 
under normal assumptions about the spatial wavefunctions showed that the exclusion principle is not violated in this approach \cite{Signal17}. 
An alternative is to use a perturbative approach involving mixing the SU(6) \textbf{56} nucleon wavefunction with higher mass \textbf{70} states. 
This approach will be examined in further work.

In figure 2 we plot the quark orbital angular momentum density, $l^{3}_{q}(x)$ 
%\bea
%l^{3}_{q}(x) = - \int d^{2} \mathbf{k}_T\;   \frac{k_{T}^{2}}{2M^{2}} h_{1T}^{\perp}(x,\ k_T)
%\eea
of the valence $u$ and $d$ quarks, calculated for a bag radius of 0.8 fm and the singlet - triplet recoil masses split by 100 MeV. 
In contrast to the usual parton distributions, $l^{3}_{u,d}(x)$ are flat over the region $0 \leq x < 0.4$ and decrease in magnitude slowly at large $x$, 
indicating that quark orbital angular momentum in the bag model is carried over a wide kinematic range, 
whereas at the bag model scale the spin dependent parton distributions are usually peaked in the valence region around $x = 0.2 - 0.3$ and quickly become small at large $x$. 
This behaviour arises because the orbital angular momentum only comes from the lower component of the relativistic quark wavefunction in equation (\ref{eq:bwf}), 
which is small at low momentum and increases to a maximum around $\kappa = 2.54$, whereas the upper component decreases over this range of momenta.
We find the valence quark orbital angular momenta are $L^{3}_{u} = +0.0211$ and $L^{3}_{d} = -0.0043$, giving a total valence contribution of 0.016. 
This small orbital angular momentum may appear to contradict the picture of Myhrer and Thomas \cite{MT88, MT08, MT10}, where the one-gluon exchange corrections to spin dependent quantities 
are dominated by diagrams involving the excitation of a $p$-wave antiquark, and results in a large fraction of the proton spin being carried by orbital angular momentum.  
However, our calculation is for valence quarks, whereas the Myhrer and Thomas result is for the sum of quarks and antiquarks, with the antiquark diagrams giving the largest contribution. 
Extending our calculations to explicitly include antiquark contributions would give further insight into the role of orbital angular momentum in the make up of the proton spin.
We note that $L^{3}$ is not gauge invariant, so this calculation is only applicable in the MIT bag model. 
However, the combination of orbital angular momentum and gluon spin $L^{3} + \Delta G$ is gauge invariant, and in the model $\Delta G = 0$, 
so this calculation does give us some insight into the portion of the proton's spin that is not carried by quarks. 

The $x$-dependent moments of transverse momentum for a given TMD are given by 
\bea
\langle k_{T}^{(1,2)} \rangle(x) = \frac{\int d^{2} \mathbf{k}_T\;  |\mathbf{k}_{T}|^{(1,2)} \phi(x, k_T)}{\int d^{2} \mathbf{k}_T\;   \phi(x, k_T)}.
\eea

In figure 3 we plot both $\langle k_T \rangle(x)$ and $\langle k_T^2  \rangle(x)$ for unpolarized valence $u$ and $d$ quarks, again calculated for $R = 0.8$ fm and 
$M_{s} - M_{t} = 100$ MeV. 
This calculation only takes into account 2-quark recoil states, and also ignores contributions from the nucleon's pion cloud. 
While we see a difference between the $u$ and $d$ quarks, this is small, 
and compatible with the experimental observation that the Gaussian widths of TMDs are flavour independent \cite{Hermes13, Compass13}. 
We find a marked $x$ dependence, in contrast to the results of \cite{AESY10}, and large values of $\langle k_T \rangle$ and $\langle k_T^2  \rangle$ 
at both low and medium $x$. 
The average transverse momentum is larger than $0.2$ GeV for $0 \leq x \leq 0.6$.
At $x = 0.3$ we find $\langle k_T^2  \rangle = (0.21, 0.19)$ GeV$^2$ for $u_{v}$ and $d_v$ respectively, compared with the values 
quoted for unpolarized quarks of $0.080$ GeV$^2$ in the light-cone constituent model calculation of \cite{BEBS09} and $0.077$ GeV$^2$ in \cite{AESY10}.
Our results are similar in magnitude to the calculation in the Nambu-Jona-Lasinio (NJL) model of \cite{MBCT12}, 
however, in the NJL model calculation $\langle k_T^2 \rangle(x)$ has only a moderate $x$-dependence, and increases at large $x$.
%While there is a difference between the $u$ and $d$ quarks, this is small, 
%and compatible with the experimental observation that the Gaussian widths of TMDs are flavour independent \cite{Hermes13, Compass13}. 

The TMDs have so far been calculated at a low momentum scale $\mu_0$ appropriate for the bag model \cite{SST91}.
In order to compare the calculated TMDs with experimental data on $\langle k_{T} \rangle$ and $\langle k^{2}_{T} \rangle$ 
we need to evolve these distributions up to experimental scales. 
Factorization of the TMDs \cite{Collins11} allows us to compare a non-singlet TMD at different scales $\mu$ and $Q$. 
Using equation (26) of reference \cite{AR11} we can write the ratio of the TMD at different scales
\bea
\frac{\tilde{F}(x, b_T, Q, \zeta_Q) }{\tilde{F}(x, b_T, \mu, \zeta_\mu) } = \frac{A(x, b_T)}{A(x, b_T)} \times \frac{B(b_T, Q, \zeta_Q) }{B(b_T, \mu, \zeta_\mu) } \times \frac{C(x, b_T, \zeta_Q) }{C(x, b_T, \zeta_\mu) }.
\eea

Here, the function $A$ is the collinear factor, which is independent of the scales $\mu$ and $Q$, and is calculated at some independent scale $\mu_{b}(b_T)$.
The function $B(b_T, \mu, \zeta)$ is calculated perturbatively in QCD and holds for all $b_T$, 
and the function $C(x, b_T, \zeta)$ describes the non-perturbative $b_T$ behaviour. 
We note the appearance of the energy cutoff scale $\zeta$ used to regulate light-cone divergences, with $\sqrt{\zeta_Q} \approx Q$ \cite{AR11, Collins11}. 
Also, this expression applies in the spatial transverse parameter ($b_T$) space, where $\tilde{F}$ is the Fourier transform of the TMD in momentum space
\bea
\tilde{F}(x, \mathbf{b}_T) & = & \int d \mathbf{k}_T \exp{(i \mathbf{k}_T \cdot \mathbf{b}_T}) \phi(x, \mathbf{k}_T)  \nonumber \\
& = & \pi \int_{p_{\mathrm{min}}}^{\infty} dp_{n} \;  \frac{p_{n}}{k_T} \phi(x, k_T) J_{0}(b_T k_T)
\eea

where we have explicitly changed integration variable to the magnitude of recoil momentum.
The ratio of non-peturbative factors only depends on the energy scales and a universal hadron independent function $g_K(b_T)$
\bea
\frac{C(x, b_T, \zeta_Q) }{C(x, b_T, \zeta_\mu) } = \exp \left[ g_K(b_T)\ln \frac{\zeta_Q}{\zeta_\mu} \right], 
\eea

where usually a quadratic form is used for $g_K = -g_2 b_T^2 /2$, giving a Gaussian model description of the TMD. 
The perturbative function $B(b_T, \mu, \zeta)$ has been calculated to first order in $\alpha_S$ \cite{AR11} and we use the NLO expression for $\alpha_S(Q)$. 
We can now take our TMD calculated in momentum space at bag scale $\mu_0$, transform to $b_T$ space, evolve up to an experimental scale $Q$, 
and finally transform back to momentum space to obtain the evolved TMD to compare with data.

In figure 4 we show the unpolarized valence up and down TMDs ($f_1$) at $x = 0.09$ for both the bag scale $\mu_0 = 0.4$ GeV and then evolved to $Q = \sqrt{2.4}$ GeV 
in both transverse momentum space and $b_T$ space.
Here we have used $g_2 = 0.68$ GeV$^{2}$, determined from a global fit \cite{LBNY03}, $\zeta_\mu = \mu^2$, $\zeta_Q = Q^2$ as usual,
and have set $b_{\mathrm{max}}$, the approximate maximum impact parameter for the perturbative domain, to 0.5 GeV$^{-1}$, 
although these calculations are not particularly sensitive to the value of $b_{\mathrm{max}}$.
%As anticipated, the shape of these distributions in impact parameter space is not well-fitted by Gaussian distributions. 

We observe that at the initial scale the calculated $\tilde{F}(x, b_T) $ become negative for $b_T$ larger than about 1.3 fm, 
which would appear to be incompatible with the interpretation of these distributions as probability distributions. 
However, it is worth noting that in QCD the positivity constraint on these distributions is only true in momentum space \cite{Collins11}. 
Additionally, these distributions are not the same as the so-called impact parameter dependent quark distributions derived from 
Generalized Parton Distributions (GPDs) \cite{Burk00, Burk02}. 
The impact parameter dependent distributions are defined in terms of the Fourier transform of a GPD $H(x, - \boldsymbol{\Delta}_{T}^{2})$, 
where $\Delta^{\mu} = p^{\prime 2} - p^{2}$ is the momentum transfer between the non-forward hadron states. 
The Fourier conjugate variable to $\boldsymbol{\Delta}_{T}$ is denoted $\mathbf{b}_{T}$ in \cite{Burk00, Burk02}, 
and is interpreted as the perpendicular distance from the centre of momentum of the target hadron. 
This is not the same as the conjugate variable to $\mathbf{k}_{T}$, which we are using for TMDs. 
Here $b_T$ is the perpendicular distance from the path of the photon to the centre of the struck quark's electric potential. 
If the two different impact parameters differ by a constant, then the shift theorem for Fourier transforms implies that the distributions will differ by a sinusoidal factor, 
and if one is positive definite, then the other is not guaranteed to always be positive. 
We hope to explore the relationship between TMDs and GPDs in further work.
%Also, this is occurring at $b_T \approx 2R$, so is most likely due to the sharp cut-off in the bag wavefunction at the bag boundary, 
%as well as the limitations of the non-relativistic Peierls-Yoccoz projection. 

Evolution of the distributions shifts them to smaller $b_T$, and the tail of the distributions at $b_T > 1.3$ fm becomes negligible. 
As expected, in momentum space evolution causes the distributions to broaden. 
For the evolved $u_v$ and $d_v$ distributions we find $\langle k^2_{T} \rangle = (0.41, 0.39)$ GeV$^{2}$ respectively, which are compatible with the value of 
$0.38 \pm 0.06$  GeV$^{2}$ found in \cite{STM10} using a Gaussian model with Hermes SIDIS data.
In addition $\langle k_{T} \rangle = (0.59, 0.57)$ GeV respectively for the unpolarized valence $u$ and $d$ quarks, which are in reasonable agreement with determinations from 
EMC and Hermes data in \cite{CEGMMS06, EGMMP05}, again derived using Gaussian models. 
We also investigated using the alternate parameters found by Konychev and Nadolsky \cite{KN06}, who found a larger value of $b_{\mathrm{max}} = 1.5$ GeV$^{-1}$ and also
$g_1 = 0.201$ GeV$^{2}$, $g_2 = 0.184$ GeV$^{2}$ and $g_3 = -0.129$.
The evolution is most sensitive to the value of $g_2$, and we find $\langle k^2_{T} \rangle = (0.38, 0.35)$ GeV$^{2}$ for the valence $u_v$ and $d_v$ distributions, 
which are also compatible with the Hermes data. 

The analysis of Anselmino \emph{et al.} \cite{ABGMP14} gives $\langle k^2_{T} \rangle = 0.57 \pm 0.08$ over the complete $x$ range of the Hermes data, with data cuts 
$Q^2 > 1.69$ GeV$^2$ and $z < 0.6$, while their fits of the Compass data give $\langle k^2_{T} \rangle = 0.61 \pm 0.20$ with similar cuts. 
These values are somewhat dependent on the data cuts, and also on the use of the Gaussian ansatz; however, given the wide range in $Q^2$ of both data sets, we would 
argue that our calculations are not incompatible with these analyses. 
We have seen that evolution of the TMDs increases their widths, so further evolution to higher $Q^2$ can give agreement with these values of $\langle k^2_{T} \rangle $.

The procedure for evolution of the TMDs could be criticised on the basis that our starting scale $\mu_0 = 0.4$ GeV is not much greater than $\Lambda_{\mathrm{QCD}} = 0.226$ GeV, 
so the evolution equations based on LO and NLO expressions may not be reliable.  
However, in the case of DIS the evolution equations appear to work quite well down to these scales, where the NLO (and NNLO) corrections do not become too large, 
and evolution of PDFs from low scales below $Q^2 = 1$ GeV$^2$ can give good agreement with the experimental PDFs \cite{SST91, SMST92, CS03, GRV98}. 
As NLO expressions for the perturbative functions $B(b_T, \mu, \zeta)$ applicable for the leading twist TMDs become available, it will be interesting to check the size of the 
corrections to LO evolution. 
We have also seen that the non-perturbative part of the evolution introduces uncertainties of the order of 5-10\%, similar to the size of differences between LO and NLO PDFs. 
Additionally, we have only calculated twist-two contributions to the distributions, and ignored higher twist contributions, which may be present in the data and could complicate 
the comparison between our calculations and the data. 

A related concern is that factorisation for TMDs requires $k^2_{T} \ll Q^{2}$ \cite{Collins11}, whereas our starting scale $\mu_0$ is similar to the size of $\langle k_{T} \rangle$ we have calculated in the bag model. 
We do not know whether this has deeper implications for the evolution of our model TMDs, however, we have seen that the increase in $\langle k^2_{T} \rangle$ as we have evolved is rather slow, 
which gives us some confidence that the corrections to our leading term are small. 

%\section*{Conclusion}
%
%Previously the bag model has been used successfully to describe the longitudinal parton distribution functions (PDFs) of the proton. 
%Here we have extended that work and presented new calculations of the twist two transverse momentum distributions in the model. 
In conclusion, we have extended earlier work using the MIT bag model to present new calculations of the twist two transverse momentum distributions in the model. 
These calculations have the correct support, and do not require any renormalization procedure. 
The distributions show strong $x$-dependence of the average transverse momentum of valence quarks. 
We have seen that longitudinal and transverse momentum components of quark momenta should not be treated as independent in the scattering process, 
which has implications for the extraction of moments of the transverse momentum distributions, and for the understanding of experimental data on transverse quark momentum.  
Using factorisation, the calculated distributions can be evolved in $Q^2$ to compare with experimental determinations of $\langle k_{T} \rangle$ and $\langle k^{2}_{T} \rangle$. 
We found that our unpolarized valence distributions gave reasonably good agreement with the current data on transverse momentum coming from experiments. 

In future work we will extend these calculations by including pion cloud contributions to the TMDs. 
This will enable us to investigate sea quark distributions as well as valence distributions, and to examine the transverse momentum dependence of the $\bar{u} - \bar{d}$ asymmetry, 
recently reported by the SeaQuest collaboration \cite{SeaQuest2021}.
%We will also generalize our calculations to the T-odd Sivers and Boer-Mulders functions and explore the relationship between TMDs and fragmentation functions.

\begin{figure}[bt]
%\centerin
\includegraphics[width=8.6cm]{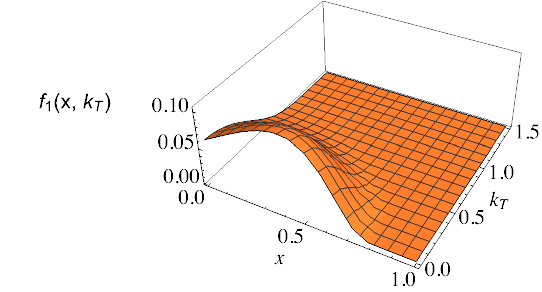}
\includegraphics[width=8.6cm]{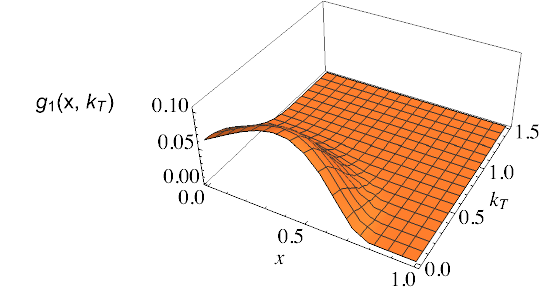}
\includegraphics[width=8.6cm]{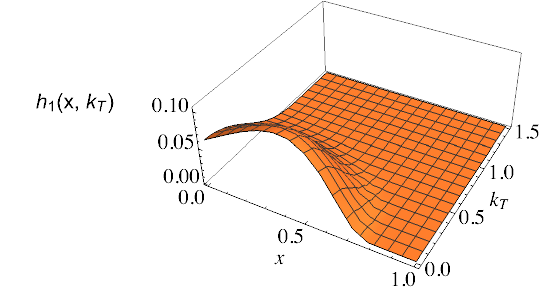}
\includegraphics[width=8.6cm]{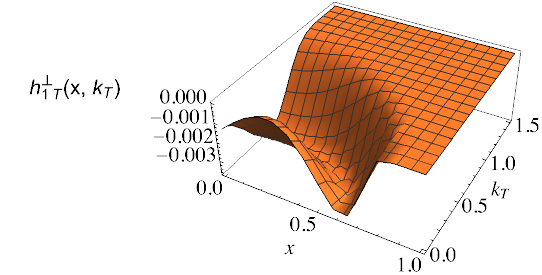}
\hfill  
\caption{Twist two transverse momentum distributions  $f_{1}(x,\ k_T)$, $g_{1}(x,\ k_T)$, $h_{1}(x,\ k_T)$ and $h_{1T}^{\perp}(x,\ k_T)$, for bag radius $R = 0.8$ fm, and where $k_T$ is in GeV.  }
\end{figure}

\begin{figure}[bt]
%\centerin
\includegraphics[width=8.6cm]{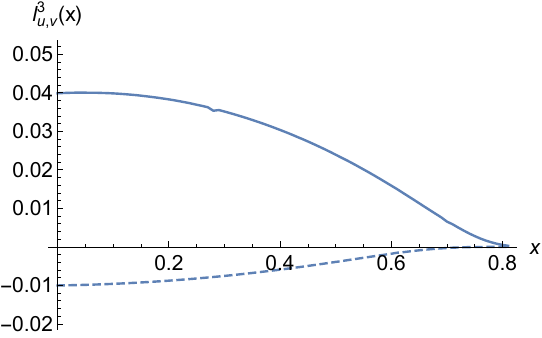}
\hfill  
\caption{Orbital angular momentum density $l^{3}_{q}(x)$ for a valence $u$ (solid line) and $d$ (dashed line) quarks at the bag model scale $\mu_{0}$.}
\end{figure}

\begin{figure}[bt]
%\centerin
\includegraphics[width=8.6cm]{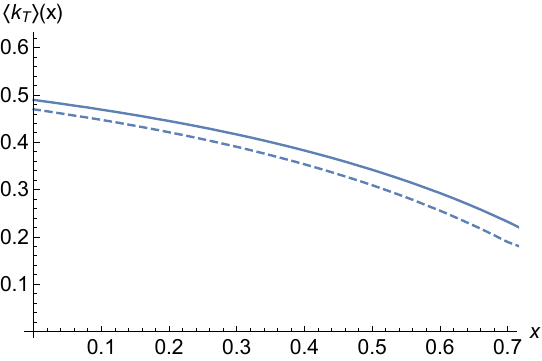}
\includegraphics[width=8.6cm]{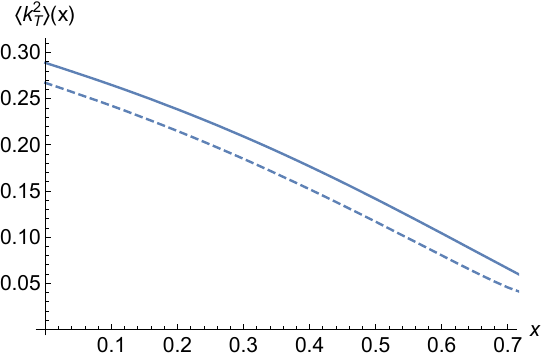}
\hfill  
\caption{Average transverse momentum $\langle k_{T} \rangle$ in GeV and average transverse momentum squared $\langle k^{2}_{T} \rangle$ in GeV$^2$
for unpolarized $u$ (solid lines) and $d$ (dashed lines) valence quarks as functions of $x$ at the bag model scale $\mu_{0}$.}
\end{figure}

\begin{figure}[bt]
%\centerin
\includegraphics[width=8.6cm]{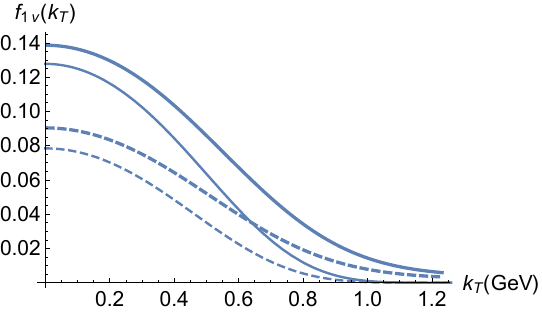}
\includegraphics[width=8.6cm]{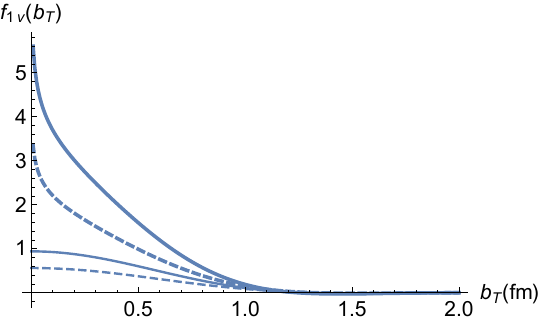}
\hfill  
\caption{Unpolarised valence quark distributions $u_v$ (solid lines) and $d_v$ (dashed lines) calculated as functions of transverse momentum $k_T$  in GeV (top) and $b_T$ in fm (bottom) at $x = 0.09$. 
Thin lines correspond to the bag scale $\mu_0 = 0.4$ GeV and thick lines to the evolved scale $Q^2 = 2.4$ GeV$^2$.}
\end{figure}

%%%%%%%%%%%%%%%%%%%%%%%%%%%%%%%%%%%%%%%%%%%%%%%%
%% BACKMATTER
%%%%%%%%%%%%%%%%%%%%%%%%%%%%%%%%%%%%%%%%%%%%%%%%

\section*{Acknowledgments}

We are grateful to Peter Schweitzer and Tony Thomas for valuable discussions on aspects of this work and for helping to correct some errors in an earlier version of this letter. 

%\section*{Corrections}



\bibliographystyle{aipproc}   % if natbib is available

%\bibliography{sample}

\begin{thebibliography}{99}

\bibitem{Collins02}
	J. C. Collins, Phys. Lett. B, 536 (2002), p. 44 

	https://doi.org/10.1016/S0370-2693(02)01819-1, https://arxiv.org/abs/hep-ph/0204004

\bibitem{Kotzinian95}
	A. Kotzinian, Nucl. Phys. B, 441 (1995), p. 234 
	
	 https://doi.org/10.1016/0550-3213(95)00098-D, https://arxiv.org/abs/hep-ph/9412283

\bibitem{Hermes13}
	A. Airapetian et al. (Hermes Collaboration), Phys. Rev. D, 87 (2013),  074029 
	
	https://doi.org/10.1103/PhysRevD.87.074029, https://arxiv.org/abs/1212.5407.

\bibitem{Compass13}
	C. Adolph et al. (Compass Collaboration), Eur. Phys. J., 73, (2013), p. 2531   
	
	https://doi.org/10.1140/epjc/s10052-013-2531-6, https://arxiv.org/abs/1305.7317 

\bibitem{Collins93}
	J. C. Collins,  Nucl. Phys. B, 396 (1993) p. 161 
	
	https://doi.org/10.1016/0550-3213(93)90262-N, https://arxiv.org/abs/hep-ph/9208213

\bibitem{Belle06}
	R. Seidl et al. (Belle collaboration),  Phys. Rev. Lett., 96 (2006), 232002 
	
	 https://doi.org/10.1103/PhysRevLett.96.232002, https://arxiv.org/abs/hep-ex/0507063

\bibitem{Belle11}
	R. Seidl et al. (Belle collaboration),  Phys. Rev. D, 99 (2019), 112006 

	https://doi.org/10.1103/PhysRevD.99.112006, https://arxiv.org/abs/1902.01552

\bibitem{AESY10}
	H. Avakian, A. V. Efremov, P. Schweitzer and F. Yuan, Phys. Rev. D 81 (2010),  074035 
	
	https://doi.org/10.1103/PhysRevD.81.074035, https://arxiv.org/abs/1001.5467

\bibitem{PCB08} 
	B. Pasquini, S. Cazzaniga and S. Boffi, Phys. Rev. D, 78 (2008), 034025 
	
	https://doi.org/10.1103/PhysRevD.78.034025, https://arxiv.org/abs/0806.2298

\bibitem{SZM09}
	J. She, J. Zhu and B-Q. Ma, Phys. Rev. D, 79 (2009), 054008   
	
	https://doi.org/10.1103/PhysRevD.79.054008, https://arxiv.org/abs/0902.3718
	
\bibitem{EPTZ09} 
	A. V. Efremov, P. Schweitzer, O. V. Teryaev and P. Zavada, Phys. Rev. D, 80 (2009) 014021   
	
	https://doi.org/10.1103/PhysRevD.80.014021, https://arxiv.org/abs/0903.3490
	
\bibitem{BCR08} 
	A. Bacchetta, F. Conti and M. Radici, Phys. Rev. D, 78, (2008) 074010   
	
	https://doi.org/10.1103/PhysRevD.78.074010, https://arxiv.org/abs/0807.0323
	
\bibitem{MBCT12} 
	H H Matevosyan, W. Bentz, I. C. Clo\"{e}t and A. W. Thomas, Phys. Rev. D, 85 (2012)  014021   
	
	https://doi.org/10.1103/PhysRevD.85.014021, https://arxiv.org/abs/1111.1740

\bibitem{Wakam09} 
	M. Wakamatsu, Phys. Rev. D, 79 (2009) 094028  
	
	https://doi.org/10.1103/PhysRevD.79.094028, https://arxiv.org/abs/0903.1886

\bibitem{Burk00} 
	M. Burkardt, Phys. Rev. D, 62 (2000) 071503  
	
	https://doi.org/10.1103/PhysRevD.62.071503, https://arxiv.org/abs/hep-ph/0105324
	
\bibitem{Burk02} 
	M. Burkardt,  Phys. Rev. D, 66 (2002) 114005  
	
	https://doi.org/10.1103/PhysRevD.66.114005, https://arxiv.org/abs/hep-ph/0209179

\bibitem{MIller07} 
	G. A. Miller,  Phys. Rev. C, 76 (2007) 065209  
	
	https://doi.org/10.1103/PhysRevC.76.065209, https://arxiv.org/abs/0708.2297  

\bibitem{Venkat11} 	
	S.Venkat, J. Arrington, G. A. Miller, and X Zhan, Phys. Rev. C, 83 (2011) 015203  
	
	https://doi.org/10.1103/PhysRevC.83.015203, https://arxiv.org/abs/1010.3629
	
\bibitem{Yuan03}
	F. Yuan, Phys. Lett. B, 575 (2003) p. 45  
	
	https://doi.org/10.1016/physletb.2003.09.052, https://arxiv.org/abs/hep-ph/0308157
	
\bibitem{AESY08}
	H. Avakian, A. V. Efremov, P. Schweitzer and F. Yuan, Phys. Rev. D, 78 (2008) 114024   
	
	https://doi.org/10.1103/PhysRevD.78.114024, https://arxiv.org/abs/0805.3355

\bibitem{ST89} 
	A. I. Signal and A. W. Thomas,  Phys. Rev. D, 40 (1989) p. 2832  
	
	https://doi.org/10.1103/PhysRevD.40.2832

\bibitem{SST91}
	A. W. Schreiber, A. I. Signal and A. W. Thomas, Phys. Rev. D, 44 (1991) p. 2653  
	
	https://doi.org/10.1103/PhysRevD.44.2653 

\bibitem{SMST92}	
	A. W. Schreiber, P. J. Mulders, A. I. Signal and A. W. Thomas, Phys. Rev. D, 45 (1992) p. 3069   
	https://doi.org/10.1103/PhysRevD.45.3069

\bibitem{GSS07}
	W. Greiner, S. Schramm and E. Stein, Quantum Chromodynamics, Third Edition, Springer-Verlag, Berlin (2007).	

\bibitem{LB80} 
	G. P. Lepage and S. J. Brodsky, Phys. Rev. D, 22 (1980) p. 2157  
	
	https://doi.org/10.1103/PhysRevD.22.2157  

\bibitem{PB85} 
	H-C, Pauli and S. J. Brodsky Phys. Rev. D, 32 (1985) p. 1993  
	
	https://doi.org/10.1103/PhysRevD.32.1993  

\bibitem{Collins11}
	J. C. Collins, Foundations of Perturbative QCD, Cambridge University Press, Cambridge (2011).
	
\bibitem{Signal97} 
	A. I. Signal, Nucl. Phys. B., 497 (1997) p. 415  
	
	https://doi.org/10.1016/S0550-3213(97)00231-9, https://arxiv.org/abs/hep-ph/9610480
	
\bibitem{PY57} 
	R. E. Peierls and J. Yoccoz, Proc. Phys. Soc, A70, (1957) p. 381 

\bibitem{LP11}
	C. Lorc\'{e} and B. Pasquini, Phys. Rev. D, 84 (2011) 034039
	
	https://doi.org/10.1103/PhysRevD.84.034039, https://arxiv.org/abs/1104.5651
	
\bibitem{LP12}
	C. Lorc\'{e} and B. Pasquini, Phys. Lett. B, 710 (2012) p. 486
	
	https://doi.org/10.1016/j.physletb.2012.03.025, https://arxiv.org/abs/1111.6069

\bibitem{RB18}
	M. Radici and A Bacchetta, Phys. Rev. Lett. 120 (2018) 192001
	
	https://doi.org/10.1103/PhysRevLett.120.192001, https://arxiv.org/abs/1802.05212

\bibitem{CHS77} 
	F. E. Close, F. Halzen and D. M. Scott, Phys. Lett. B, 68 (1977) p. 447 
	
	https://doi.org/10.1016/0370-2693(77)90466-X

\bibitem{DS77}
	A. C. Davis and E. J. Squires, Phys. Lett. B, 69 (1977) p. 249  
	
	 https://doi.org/10.1016/0370-2693(77)90655-4

\bibitem{MT88}
	F. Myhrer and A. W. Thomas,Phys. Rev. D, 38 (1988) 1633   
	
	https://doi.org/10.1103/PhysRevD.38.1633

\bibitem{MT08}
	F. Myhrer and A. W. Thomas, Phys. Lett. B, 663 (2008) p. 302   
	
	https://doi.org/10.1016/j.physletb.2008.04.034, https://arxiv.org/abs/0709.4067 
	
\bibitem{MT10}	
	F. Myhrer and A. W. Thomas, J. Phys. G, 37 (2010) 023101 
	
	 http://doi.org/10.1088/0954-3899/37/2/023101, https://arxiv.org/abs/0911.1974

\bibitem{ABGMP14}
	M. Anselmino, M. Boglione, J. O. Gonzalez H, S. Melis and A. Prokudin, JHEP,  04 (2014) 005 
	https://doi.org/10.1007/JHEP04(2014)005 https://arxiv.org/abs/1312.62617a

\bibitem{CT88}
	F. E. Close and A. W. Thomas, Phys. Lett. B, 212 (1988) p. 227  
	
	https://doi.org/10.1016/0370-2693(88)90530-8

\bibitem{Isgur99}
	N. Isgur, Phys. Rev. D, 59 (1999) 034013  
	
	https://doi.org/10.1103/PhysRevD.59.034013, https://arxiv.org/abs/hep-ph/9809255
	
\bibitem{Signal17}
	A. I. Signal,  Phys. Rev. D, 95 (2017) 114010   
	
	https://doi.org/10.1103/PhysRevD.95.114010, https://arxiv.org/abs/1702.05152

\bibitem{BEBS09}
	S. Boffi, A. V. Efremov, B. Pasquini and P. Schweitzer, Phys. Rev. D, 79 (2009) 094012  
	
	[https://doi.org/10.1103/PhysRevD.79.094012, https://arxiv.org/abs/0903.1271].

\bibitem{AR11}
	S. M. Aybat and T. C. Rogers, Phys. Rev. D, 83 (2011) 114042  
	
	 https://doi.org/10.1103/PhysRevD.83.114042, https://arxiv.org/abs/1101.5057

\bibitem{LBNY03} 
	F. Landry, R. Brock, P. M. Nadolsky and C. P. Yuan,  Phys. Rev. D, 67 (2003) 073016  
	
	https://doi.org/10.1103/PhysRevD.67.073016, https://arxiv.org/abs/hep-ph/0212159
	
\bibitem{STM10} 
	P. Schweitzer, T. Teckentrup, and A. Metz, Phys. Rev. D, 81 (2010) 094019  
	
	https://doi.org/10.1103/PhysRevD.81.094019, https://arxiv.org/abs/1003.2190	
	
\bibitem{CEGMMS06}
	J. C. Collins, A. V. Efremov, K. Goeke, S. Menzel, A. Metz and P. Schweitzer, Phys. Rev. D, 73 (2006) 014021 
	
	https://doi.org/10.1103/PhysRevD.73.014021, https://arxiv.org/abs/hep-ph/0509076 

\bibitem{CEG06}	
	J. C. Collins et al. Phys. Rev. D, 73 (2006) 094023   
	
	https://doi.org/10.1103/PhysRevD.73.094023, https://arxiv.org/abs/hep-ph/0511272

\bibitem{EGMMP05}
	A. V. Efremov, K. Goeke, S. Menzel, A. Metz and P. Schweitzer, Phys. Lett. B, 612 (2005) 233  
	
	https://doi.org/10.1016/j.physletb.2005.03.010, https://arxiv.org/abs/hep-ph/0412353
	
\bibitem{KN06}
	A. V. Konychev and P, M. Nadolsky, Phys. Lett. B, 633 (2006) 710 
	
	https://doi.org/10.1016/j.physletb.2005.12.063, https://arxiv.org/abs/hep-ph/0506225
	
\bibitem{CS03}
	F. G. Cao and A. I. Signal, Phys. Rev. D,  68 (2003) 074002  
	
	https://doi.org/10.1103/PhysRevD.68.074002, https://arxiv.org/abs/hep-ph/0306033

\bibitem{GRV98} M. Gl\"{u}ck, E. Reya and A. Vogt Eur. Phys. J., 5 (1998) p. 461

	https://doi.org/10.1007/s100529800978, https://arxiv.org/abs/hep-ph/9806404
	
\bibitem{SeaQuest2021}
	J. Dove et al. (SeaQuest collaboration), Nature, 590 (2021) p. 561  
	
	https://doi.org/10.1038/s41586-021-03282-z, https://arxiv.org/abs/2103.04024


\end{thebibliography}

%  \typeout{******************************************}
 % \typeout{** Please run "bibtex \jobname" to optain}
%  \typeout{** the bibliography and then re-run LaTeX}
%  \typeout{** twice to fix the references!}
%  \typeout{******************************************}

\end{document}